\documentstyle[aps,psfig,multicol]{revtex}

\begin{document}
\draft 

\title{ Coherence length of neutron superfluids}
 
\author{G.\ Lazzari$^{a,b}$, F.\ V.\ De Blasio$^{c}$, 
M.\ Hjorth-Jensen$^{c}$, \O.\ Elgar\o y$^d$ and L.\ Engvik$^d$ }

\address{ $^a$ European Centre for Theoretical Studies 
in Nuclear Physics and Related Areas, Trento, Italy}
\address{$^b$ Dipartimento di Fisica,
Universit\`a di Milano, Via Celoria 16, I-20133 Milano, Italy}
\address{$^c$ NORDITA, Blegdamsvej 17, DK-2100 Copenhagen \O, Denmark}
\address{$^d$ Department of Physics, University of Oslo, N-0316 Oslo, Norway}

\maketitle
 
\begin{abstract}

The coherence length of superfluid neutron matter 
is calculated  from the microscopic BCS wavefunction of a Cooper  
pair in momentum space making use of the 
 Bonn meson-exchange potential. 
We find that the coherence length is  proportional to the 
Fermi momentum-to pairing gap ratio, in good agreement with 
simple  estimates used in the literature, and we establish 
the appropriate fitting constants using our numerical data. 
Our calculations can be applied to  the problem of inhomogeneous superfluidity 
of hadronic matter in the crust of a neutron star.
 
\end{abstract}


\begin{multicols}{2}

Calculations based on BCS theory with 
phenomenological nucleon-nucleon (NN) forces 
indicate that neutron matter is superfluid in a wide region of 
densities and temperatures.     
In particular, at 
the densities corresponding to the interior of a neutron star crust 
($ 4\times 10^{11}\;{\rm gcm}^{-3} < \rho  < 
10^{14}\;{\rm  gcm}^{-3} $) 
neutrons couple in the singlet isotropic channel $^1 S_0$  
while at larger 
densities they pair in the $^3P_2$ state \cite{elg96}. 
Although many investigations have been devoted to the 
superfluidity and superconductivity    
of neutron, $\beta$-stable and nuclear matter, little attention 
has been paid to possible effects of inhomogeneities 
in hadronic superfluids or superconductors. In fact 
the interior of neutron stars, the subject of most of such 
calculations, is often referred to as the 
only existing example in nature of infinite 
superfluid neutron or $\beta$-stable 
matter, since in an atomic nucleus the wavefunctions of 
the Cooper pairs  are limited in extension by the potential well. 
On the other hand, 
superfluidity in a neutron star crust represents a 
case intermediate between the nucleus and the idealized infinite system, 
since  superfluid neutrons in the 
inner crust occupy a region  where a lattice of nuclei 
creates strong inhomogeneities in the medium.   
From the point of view of astrophysical observations,       
probably  
most of the visible effects of the presence of superfluids 
in a neutron star are due to phenomena  in the crust \cite{petra}.

An important  length scale of the neutron superfluid   
is the coherence length.
From a microscopic point of view the 
coherence length represents 
the squared mean distance   of two paired particles 
(a Cooper pair of neutrons)  
on top of the Fermi surface. 
The magnitude of this quantity affects several of the physical properties 
of a neutron star crust. 
First of all, neutrons paired in a singlet state form quantized vortices 
induced by the rotational 
state of the star. These   can pin to the nuclei present in the 
crust, possibly leading to the observed sudden release of angular momentum 
known as pulsar glitches. The magnitude of the pinning force 
depends on the size of the vortex cores, 
which is equal to the coherence length of the neutron superfluid. A second 
question is how properties of the neutron 
superfluid change due to  the inhomogeneous 
environment of a neutron star crust, a problem related to the average 
thermodynamical property of neutron  matter \cite{petra,noi}.  
In the inner crust, depending on density, 
nuclei of different shapes and sizes are 
present. At a density 
of $ \sim 10^{14}~{\rm gcm}^{-3}$, spherical nuclei cease to be 
energetically favored and are replaced first by   
cylindrical nuclei, then slabs to end up   
with holes, where  the roles of protons and neutrons are 
exchanged, see Refs.\  \cite{petra,lorenz} for further details. 
Only at higher densities, corresponding to what is called the core 
of the star, do nuclei merge into the uniform medium.  
The fact that  neutron superfluidity in neutron star matter  
is actually a  problem of 
inhomogeneous superfluidity 
in hadronic matter has been noticed  quite recently \cite{petra,noi}.
According to Anderson's theorem \cite{degennes} 
the electron density of states in a superconductor  
is changed very little from a pure metal to an alloy of 
similar chemical properties. The physical situation we are examining here 
is quite different, since both the average density of states and the 
effective neutron-neutron matrix elements are changed compared to the 
uniform case when one considers the presence of nuclei.  

The typical dimension of nuclei 
in the inner crust of a neutron star 
is $R_{N}\approx 4 - 6\;{\rm fm}$.  This number is, in an appropriate range 
of densities, comparable to the coherence length $\xi$ as 
estimated from existing BCS calculations.
  If Anderson's theorem holds also for neutron star matter, 
(limit where $R_{N}\ll\xi$)   
there will be no appreciable variations in the superfluid properties 
induced by the nuclei. On the other hand, if $R_{N}\gg\xi$ the superfluid 
will change its properties locally. This limit has been investigated  
in a recent series of papers \cite{noi}  where it was found that some  
thermodynamical properties like e.g. the neutron specific heat may change 
by a very large amount. 
Unfortunately, the situation 
is complicated by the fact that $R_{N}$ and $\xi$ are of 
the same order of magnitude.

Clearly, the coherence length represents a critical parameter by which
one can 
establish the behavior of an inhomogeneous superfluid. 
It sets the scale for the possible spatial variation of the pairing 
properties of the system, and thus plays a role if some 
inhomogeneities are present in the system at a length scale 
comparable to it. 

A simple estimate of the coherence length is obtained by assuming  
constant matrix elements between particle states within a shell 
centered at the Fermi momentum and zero outside 
(see for example Ref. \cite{degennes}).  
This gives
\begin{equation}
    \xi \approx K {k_F \over \Delta_F} 
    \label{eq:eq1}
\end{equation}
where   $k_{F}$ 
is the Fermi momentum
and $\Delta_F$ is the 
pairing gap at the Fermi momentum.   
The value of the constant $K$   depends on the approximations used, 
and different values have been reported in the literature. 
However, the estimate (\ref{eq:eq1})  
does not consider thoroughly 
the matrix elements of the particle-particle interactions, and in principle   
can be too rough if one wants to calculate the 
coherence length at all values of the density. 
In view     
of the important astrophysical considerations discussed above, 
it is desirable  to  calculate  the coherence length
 from a microscopic study of the wavefunction of a Cooper pair. 
The aim of this  Brief Report is thus twofold: 
firstly we check the validity of the simple estimate (\ref{eq:eq1}) and 
search for possible deviations or fluctuations from its average behavior. 
Secondly, in those cases where (\ref{eq:eq1}) is well reproduced, we determine 
the best value of $K$ and check the dependence of that formula 
on the neutron interaction in the particle-particle channel. To this 
end, we shall make use of the Bonn A potential.  
This meson-exchange potential is defined by the parameters of 
Table A.1 in Ref. \cite{bonn}.

Our microscopic calculations of coherence lengths have been carried out 
as follows.  Let  $\phi({\bf r})$ be the wavefunction of the 
relative motion of the two neutrons in a Cooper pair, ${\bf r}$ 
being the relative coordinate of the two particles.   
The coherence length $\xi$ is given by 
\begin{eqnarray}
     \xi^2&=&{ \int d^3 x |\phi({\bf r})|^2 r^2 \over 
     \int d^3 x |\phi({\bf r})|^2 }= 
     {\int d^3 k |{\partial \over \partial {\bf k}}
     \chi({\bf k})|^2 \over \int d^3 k |\chi({\bf k})|^2 } \nonumber \\ 
     &=& {\int_0^{\infty} dk k^2 |\partial\chi(k)/\partial k |^2 \over 
     \int_{0}^{\infty} dk k^2 |\chi(k)|^2 } 
     \label{eq:eq2}
\end{eqnarray} 
where in  the second equation the expectation value 
is calculated in momentum space,    
 $\chi({\bf k})$ being the Fourier transform of $\phi({\bf r})$. 
Eq.\ (\ref{eq:eq2}) is particularly suited for numerical computation,  
since the  BCS equations for a uniform system are solved in momentum space. 
The wavefunction of the Cooper pair in momentum space is 
given by (apart from an unimportant normalization constant)  
\begin{equation}
     \chi({ k})={\Delta(k) \over E(k)}
     \label{eq:eq3} 
\end{equation} 
where $\Delta(k)$ is the $k$-dependent pairing gap, 
while $E(k)={\sqrt{ \Delta^{2}(k)
+(\epsilon(k)-\mu)^2 }}$ is the quasi-particle 
energy and $\epsilon(k)$, $\mu$ are the neutron single-particle 
energy and chemical potential respectively. Notice that due to 
the isotropic nature of $^{1}S_{0}$ pairing all 
quantities depend only on the magnitude of the momentum. 

We calculate the  pairing gaps and wavefunctions in momentum space 
as functions of the momentum from the BCS gap equation \cite{elg96} 
\begin{equation}
      \Delta(p)=-{1\over 2} 
      \int { d^3 ~ k \over (2\pi)^3 } 
      \tilde{v}(p,k){\Delta(k) \over E(k)}
      \label{eq:eq4}
\end{equation}
for different values of the Fermi momentum.   
Here $\tilde{v}(p,k)$ 
are the matrix elements of the $^{1}S_{0}$ NN interaction, 
obtained from the Bonn A potential, following the procedure
outlined in Ref.\ \cite{elg96}.
The single-particle energies 
$\epsilon(k)$ were obtained from Brueckner-Hartree-Fock calculations 
with the same potential, see Ref.\ \cite{elg96,hko95} for details.  
The chemical potential $\mu$ was set equal to the Fermi energy 
$\epsilon(k_{F})$.  
Notice that for our purposes the pairing gap has to be calculated  at all 
wavenumbers  and not only at the Fermi surface, as in 
the simple formula (\ref{eq:eq1}).  
The wavefunctions $\chi(k)$ in momentum space for four different values 
of the Fermi momentum ($k_f=$0.1, 0.4, 0.8 and 1.2 fm$^{-1}$) 
$k_{F}$ are shown in Fig.\ 1. The square of the  
derivatives of the respective wavefunctions 
are displayed in Fig.\ 2. It is seen that the spread 
of the wavefunction in momentum space $<{\bf k}^{2}>$ varies considerably 
according to the value of the Fermi wavenumber 
 $k_{F}$. The distribution in momentum space is particularly wide 
for intermediate values of $k_{F}$ and, since we have through the
uncertainty relation that
$<{\bf k}^2 > = \xi^2 \approx 1/ <{\bf r}^{2}>$, 
a smaller value of the coherence length for these values is expected. 
Due to the limited, but significant variation of the
the pairing gap $\Delta(k)$   within a single peak (each peak occurs 
at the Fermi momentum), the wavefunction is 
distorted with respect to a simple gaussian and therefore 
deviations from the simple relation (\ref{eq:eq1}) for a gaussian 
distribution should be expected. This becomes evident by looking at the 
 squared derivative of the wavefunction 
$ (\partial \chi(k) / \partial k )^2 $ which shows two unequal peaks.  

In Fig.\ 3 we present 
the squared wavefunctions $r^2 |\phi(r)|^2 $  for the same 
values of the Fermi momentum as in Figs.\ 1 and 2. 
The wavefunctions oscillate with a 
wavelength of the order $k_{F}^{-1}$ and extend for quite long 
distances compared to the range of nuclear forces (a few fermis 
at most). Not surprisingly, this behavior is 
due to the relatively small value of 
the energy of the pair compared to the    
Fermi energy (a comparison can be made 
with another weakly bound state with a very spread 
wavefunction, the deuteron \cite{weis}).   

In Fig.\ 4 we show our microscopic calculations of the coherence length 
for pure neutron matter performed using Eqs.\
(\ref{eq:eq2}-\ref{eq:eq4}).  
The microscopically calculated points, given by the solid line, 
can be fitted by a
function linear in the parameter 
 $k_{F}/\Delta_{F}$, where $\Delta_{F}$ is the 
pairing gap at the Fermi surface. We find as a best fit, see Ref.\
\cite{anderson} for further details on the form for $\xi$,
\begin{equation}
      \xi= \theta \frac{\hbar^2}{2m}\frac{k_F}{\Delta_{F}}x
      \label{eq:eq5}
\end{equation}
where  $x=m/m^*$ is the ratio between the  bare and the 
effective mass, and the dimensionless 
parameter $\theta $ is found by a direct fit to our data  
to be $\theta=0.814$  (for $x=1$) and $\theta=0.836$ (with 
the correct density dependent value of $x$ from the miscroscopic
calculation). The choice $x=1$ is commonly used in 
the literature.

The simple linear behavior of the momentum-to gap ratio
is in very good agreement with Eq.\ (\ref{eq:eq1})  
and shows that a Cooper pair represents a packet 
having a nearly well-defined relative momentum. 
As seen from Fig.\ 4, 
the use of $x=1$ (effective mass equal to bare mass) does not change 
the picture very much, as is evident from the fact that at e.g., the 
relatively high Fermi momentum of $k_F = 1.0\;{\rm fm}^{-1}$ the 
effective mass differs from the bare one by less than $4\%$. 
However, deviations from Eq.\ (\ref{eq:eq1}) are found, as can be
seen in Fig.\ 4, especially at higher densities.   
In Fig.\ 4 we have also plotted the values of the coherence length 
from an approximation often used in neutron star studies, 
namely the choice of 
a coefficient $\theta=2/\pi=0.637 $ in Eq.\ (\ref{eq:eq5}) 
\cite{anderson}. 
Our values were on the average more than $ 20 \% $ larger than 
the results obtained with this choice. It is therefore
interesting to see that our microscopic calculation of the 
gap and single-particle properties through a complicated
many-body scheme, yields a qualitative similar result as that
of Eq.\ (1), derived originally in a solid state context \cite{degennes}.

A relevant point is whether  
the value of the coefficient $\theta$ depends on the interaction chosen. 
Besides, it might be that the good agreement with Eq.\ (\ref{eq:eq5})  
gets worse with other interactions. To partially answer these questions, we 
repeated the same calculations making use of an effective force, the 
Gogny force, which pairing properties have been investigated 
in several works \cite{noi,gogny}.  
The Gogny force is an effective interaction fitted to reproduce
various nuclear data, and contains therefore
effects 
of nucleon-nucleon correlations not included in a bare
nucleon-nucleon interaction. It 
is therefore both qualitatively and quantitatively different from 
a bare NN interaction like the Bonn potential.    
We find that with the D1 parameterization 
of the Gogny force  a very good fit 
can be made for Eq.\  (3) with a coefficient $\theta=0.815$, 
very close to the value found with Bonn A.  
This near-equality of the coefficient $\theta$ for the two 
different interactions is not predictable on the basis of 
simple arguments.  However, it ought to be stressed that we have
not included so-called polarization effects in the
calculations of the pairing gaps of Ref.\ \cite{elg96}.
Such effects are expected to reduce by at least a factor of
two the value of the pairing gap, see Ref.\ \cite{schulze96}.
How these many-body effects will change
the coherence length remains to be investigated. However, using the 
weak-coupling approach of Eq.\ (\ref{eq:eq1})
for the coherence length, one sees that $\xi$ may be twice as large. 

To conclude, we calculated microscopically the coherence 
length of superfluid neutron matter. Eq.\ (\ref{eq:eq5}) with 
the nearly interaction-independent 
coefficient $\theta=0.83$ represents the main result of 
our calculations. 
The data reported in this brief report may be useful 
in quantitative calculations of the superfluid properties  
of neutron star crusts. Moreover, 
as can be seen from Fig.\ 4, the coherence
length is for several Fermi momenta larger than the typical
size of known nuclei. This poses several constraints on the use
of local density approximations in the study of neutron star
crust properties \cite{petra,noi}.

\acknowledgments{
One of us (F.D.B.) acknowledges partial support from "Fondazione 
Angelo della Riccia" in Firenze, Italy. 
This work has received support from The Research Council of Norway 
(NFR) (Programme for Supercomputing) through a grant of computing time.}

\end{multicols}

\clearpage

\begin{figure}
\setlength{\unitlength}{0.1bp}
\special{!
/gnudict 40 dict def
gnudict begin
/Color false def
/Solid false def
/gnulinewidth 5.000 def
/vshift -33 def
/dl {10 mul} def
/hpt 31.5 def
/vpt 31.5 def
/M {moveto} bind def
/L {lineto} bind def
/R {rmoveto} bind def
/V {rlineto} bind def
/vpt2 vpt 2 mul def
/hpt2 hpt 2 mul def
/Lshow { currentpoint stroke M
  0 vshift R show } def
/Rshow { currentpoint stroke M
  dup stringwidth pop neg vshift R show } def
/Cshow { currentpoint stroke M
  dup stringwidth pop -2 div vshift R show } def
/DL { Color {setrgbcolor Solid {pop []} if 0 setdash }
 {pop pop pop Solid {pop []} if 0 setdash} ifelse } def
/BL { stroke gnulinewidth 2 mul setlinewidth } def
/AL { stroke gnulinewidth 2 div setlinewidth } def
/PL { stroke gnulinewidth setlinewidth } def
/LTb { BL [] 0 0 0 DL } def
/LTa { AL [1 dl 2 dl] 0 setdash 0 0 0 setrgbcolor } def
/LT0 { PL [] 0 1 0 DL } def
/LT1 { PL [4 dl 2 dl] 0 0 1 DL } def
/LT2 { PL [2 dl 3 dl] 1 0 0 DL } def
/LT3 { PL [1 dl 1.5 dl] 1 0 1 DL } def
/LT4 { PL [5 dl 2 dl 1 dl 2 dl] 0 1 1 DL } def
/LT5 { PL [4 dl 3 dl 1 dl 3 dl] 1 1 0 DL } def
/LT6 { PL [2 dl 2 dl 2 dl 4 dl] 0 0 0 DL } def
/LT7 { PL [2 dl 2 dl 2 dl 2 dl 2 dl 4 dl] 1 0.3 0 DL } def
/LT8 { PL [2 dl 2 dl 2 dl 2 dl 2 dl 2 dl 2 dl 4 dl] 0.5 0.5 0.5 DL } def
/P { stroke [] 0 setdash
  currentlinewidth 2 div sub M
  0 currentlinewidth V stroke } def
/D { stroke [] 0 setdash 2 copy vpt add M
  hpt neg vpt neg V hpt vpt neg V
  hpt vpt V hpt neg vpt V closepath stroke
  P } def
/A { stroke [] 0 setdash vpt sub M 0 vpt2 V
  currentpoint stroke M
  hpt neg vpt neg R hpt2 0 V stroke
  } def
/B { stroke [] 0 setdash 2 copy exch hpt sub exch vpt add M
  0 vpt2 neg V hpt2 0 V 0 vpt2 V
  hpt2 neg 0 V closepath stroke
  P } def
/C { stroke [] 0 setdash exch hpt sub exch vpt add M
  hpt2 vpt2 neg V currentpoint stroke M
  hpt2 neg 0 R hpt2 vpt2 V stroke } def
/T { stroke [] 0 setdash 2 copy vpt 1.12 mul add M
  hpt neg vpt -1.62 mul V
  hpt 2 mul 0 V
  hpt neg vpt 1.62 mul V closepath stroke
  P  } def
/S { 2 copy A C} def
end
}
\begin{picture}(3600,2160)(0,0)
\special{"
gnudict begin
gsave
50 50 translate
0.100 0.100 scale
0 setgray
/Helvetica findfont 100 scalefont setfont
newpath
-500.000000 -500.000000 translate
LTa
600 251 M
2817 0 V
600 251 M
0 1858 V
LTb
600 251 M
63 0 V
2754 0 R
-63 0 V
600 589 M
63 0 V
2754 0 R
-63 0 V
600 927 M
63 0 V
2754 0 R
-63 0 V
600 1264 M
63 0 V
2754 0 R
-63 0 V
600 1602 M
63 0 V
2754 0 R
-63 0 V
600 1940 M
63 0 V
2754 0 R
-63 0 V
600 251 M
0 63 V
0 1795 R
0 -63 V
1240 251 M
0 63 V
0 1795 R
0 -63 V
1880 251 M
0 63 V
0 1795 R
0 -63 V
2521 251 M
0 63 V
0 1795 R
0 -63 V
3161 251 M
0 63 V
0 1795 R
0 -63 V
600 251 M
2817 0 V
0 1858 V
-2817 0 V
600 251 L
LT0
3114 1946 M
180 0 V
600 972 M
1 0 V
1 0 V
1 0 V
1 0 V
1 1 V
1 0 V
1 1 V
2 1 V
1 1 V
2 1 V
1 2 V
2 1 V
2 3 V
2 3 V
2 3 V
3 4 V
2 4 V
2 5 V
3 6 V
3 7 V
2 8 V
3 9 V
3 10 V
3 11 V
3 13 V
3 14 V
4 16 V
3 18 V
3 20 V
4 23 V
3 25 V
4 29 V
3 31 V
4 36 V
4 40 V
4 44 V
3 50 V
4 55 V
4 60 V
4 65 V
4 69 V
4 71 V
4 69 V
4 62 V
4 47 V
4 26 V
4 -3 V
4 -32 V
4 -61 V
4 -82 V
4 -94 V
4 -99 V
4 -99 V
3 -93 V
4 -86 V
4 -78 V
4 -71 V
4 -64 V
3 -56 V
4 -51 V
4 -45 V
3 -41 V
4 -36 V
3 -33 V
4 -30 V
3 -26 V
3 -24 V
3 -22 V
3 -20 V
3 -18 V
3 -17 V
3 -15 V
3 -14 V
3 -12 V
2 -12 V
3 -10 V
2 -10 V
2 -9 V
2 -8 V
2 -7 V
2 -7 V
2 -6 V
2 -6 V
1 -5 V
2 -4 V
1 -5 V
2 -3 V
1 -4 V
1 -2 V
1 -3 V
0 -2 V
1 -2 V
1 -1 V
0 -1 V
0 -1 V
2 -6 V
9 -21 V
16 -31 V
22 -34 V
29 -33 V
36 -28 V
42 -23 V
49 -18 V
55 -14 V
60 -11 V
67 -8 V
72 -6 V
77 -5 V
82 -4 V
86 -3 V
90 -2 V
95 -2 V
98 -1 V
101 -1 V
104 -1 V
106 -1 V
108 0 V
110 -1 V
110 0 V
112 0 V
111 0 V
111 -1 V
111 0 V
110 0 V
108 0 V
59 0 V
LT1
3114 1846 M
180 0 V
601 998 M
1 0 V
2 0 V
1 0 V
2 1 V
2 0 V
3 0 V
3 0 V
3 0 V
3 1 V
4 0 V
4 1 V
4 0 V
5 1 V
5 1 V
5 1 V
5 2 V
6 1 V
6 2 V
6 2 V
6 2 V
7 3 V
7 2 V
7 3 V
7 4 V
8 4 V
8 4 V
8 5 V
8 5 V
8 6 V
8 6 V
9 6 V
9 8 V
9 8 V
9 9 V
9 9 V
9 10 V
10 11 V
9 12 V
10 13 V
9 14 V
10 15 V
10 16 V
10 17 V
10 19 V
10 20 V
10 21 V
10 23 V
10 25 V
10 26 V
10 28 V
10 30 V
10 32 V
9 33 V
10 36 V
10 37 V
10 38 V
9 40 V
10 41 V
9 41 V
10 41 V
9 40 V
9 38 V
9 35 V
8 31 V
9 26 V
9 19 V
8 12 V
8 4 V
8 -3 V
8 -11 V
7 -18 V
7 -24 V
7 -29 V
7 -34 V
7 -35 V
6 -38 V
7 -38 V
5 -39 V
6 -37 V
5 -37 V
6 -34 V
4 -33 V
5 -31 V
4 -29 V
4 -27 V
4 -24 V
4 -22 V
3 -20 V
3 -18 V
2 -16 V
2 -14 V
2 -12 V
2 -9 V
1 -8 V
1 -6 V
1 -5 V
0 -2 V
2 -11 V
8 -44 V
14 -74 V
20 -95 V
26 -106 V
32 -105 V
38 -99 V
43 -87 V
49 -74 V
54 -63 V
59 -52 V
65 -42 V
68 -35 V
73 -28 V
78 -23 V
80 -19 V
85 -15 V
87 -13 V
91 -10 V
92 -9 V
95 -7 V
97 -5 V
97 -5 V
99 -4 V
100 -3 V
99 -2 V
100 -2 V
98 -2 V
98 -1 V
5 0 V
LT2
3114 1746 M
180 0 V
601 721 M
1 0 V
2 0 V
2 0 V
3 0 V
4 0 V
4 0 V
4 0 V
6 0 V
5 0 V
6 0 V
7 1 V
7 0 V
8 0 V
9 1 V
8 0 V
9 1 V
10 0 V
10 1 V
11 1 V
11 1 V
12 1 V
12 1 V
12 1 V
13 2 V
13 1 V
14 2 V
13 2 V
15 2 V
14 3 V
15 3 V
15 3 V
16 3 V
16 3 V
16 4 V
16 4 V
17 5 V
17 5 V
17 5 V
17 6 V
17 6 V
17 6 V
18 8 V
18 8 V
17 8 V
18 9 V
18 10 V
18 11 V
18 12 V
18 12 V
18 14 V
18 14 V
18 16 V
18 18 V
18 19 V
17 20 V
18 23 V
18 24 V
17 27 V
17 29 V
17 32 V
17 34 V
16 38 V
17 42 V
16 45 V
15 49 V
16 53 V
15 57 V
15 61 V
15 65 V
14 66 V
14 67 V
14 66 V
13 61 V
12 52 V
13 41 V
12 25 V
11 9 V
11 -8 V
11 -24 V
10 -37 V
10 -45 V
9 -52 V
9 -54 V
8 -54 V
8 -52 V
7 -50 V
7 -46 V
6 -42 V
6 -37 V
5 -34 V
5 -30 V
4 -25 V
3 -22 V
3 -17 V
3 -15 V
1 -10 V
2 -8 V
0 -4 V
2 -10 V
6 -37 V
12 -62 V
17 -81 V
22 -93 V
27 -96 V
31 -92 V
37 -85 V
41 -76 V
45 -65 V
50 -56 V
54 -47 V
58 -40 V
61 -33 V
65 -28 V
68 -23 V
71 -20 V
73 -16 V
76 -14 V
78 -11 V
79 -9 V
81 -8 V
83 -7 V
83 -5 V
83 -4 V
84 -4 V
25 -1 V
LT3
3114 1646 M
180 0 V
601 401 M
2 0 V
3 0 V
3 0 V
4 0 V
5 0 V
6 0 V
7 0 V
7 0 V
8 0 V
9 0 V
10 0 V
11 0 V
11 0 V
12 0 V
13 0 V
13 0 V
14 1 V
15 0 V
15 0 V
16 0 V
17 0 V
17 1 V
18 0 V
18 0 V
19 1 V
20 0 V
20 0 V
21 1 V
21 0 V
21 1 V
22 1 V
23 0 V
22 1 V
24 1 V
23 1 V
24 1 V
24 1 V
25 1 V
24 1 V
25 1 V
26 2 V
25 1 V
26 2 V
25 2 V
26 2 V
26 2 V
26 2 V
26 3 V
26 3 V
26 3 V
26 3 V
26 4 V
26 4 V
25 4 V
26 4 V
25 6 V
25 5 V
25 7 V
25 7 V
25 7 V
24 9 V
24 10 V
23 11 V
24 12 V
22 14 V
23 16 V
22 19 V
21 21 V
21 25 V
21 30 V
20 36 V
20 44 V
19 54 V
18 68 V
18 87 V
17 113 V
17 151 V
16 198 V
15 242 V
15 225 V
14 62 V
13 -153 V
13 -237 V
12 -213 V
11 -167 V
11 -125 V
9 -94 V
9 -73 V
9 -56 V
7 -44 V
7 -35 V
5 -28 V
6 -22 V
4 -17 V
3 -13 V
3 -10 V
2 -6 V
1 -4 V
1 -5 V
5 -17 V
10 -28 V
13 -34 V
18 -38 V
22 -36 V
26 -35 V
29 -31 V
33 -27 V
37 -24 V
40 -20 V
44 -17 V
47 -15 V
49 -12 V
53 -10 V
55 -9 V
57 -7 V
60 -7 V
61 -5 V
63 -4 V
65 -4 V
49 -2 V
stroke
grestore
end
showpage
}
\put(3054,1646){\makebox(0,0)[r]{$k_F=1.2$ fm$^{-1}$}}
\put(3054,1746){\makebox(0,0)[r]{$k_F=0.8$ fm$^{-1}$}}
\put(3054,1846){\makebox(0,0)[r]{$k_F=0.4$ fm$^{-1}$}}
\put(3054,1946){\makebox(0,0)[r]{$k_F=0.1$ fm$^{-1}$}}
\put(2008,51){\makebox(0,0){$k$ (fm$^{-1}$)}}
\put(100,1180){%
\special{ps: gsave currentpoint currentpoint translate
270 rotate neg exch neg exch translate}%
\makebox(0,0)[b]{\shortstack{$\chi (k)$}}%
\special{ps: currentpoint grestore moveto}%
}
\put(3161,151){\makebox(0,0){2}}
\put(2521,151){\makebox(0,0){1.5}}
\put(1880,151){\makebox(0,0){1}}
\put(1240,151){\makebox(0,0){0.5}}
\put(600,151){\makebox(0,0){0}}
\put(540,1940){\makebox(0,0)[r]{1}}
\put(540,1602){\makebox(0,0)[r]{0.8}}
\put(540,1264){\makebox(0,0)[r]{0.6}}
\put(540,927){\makebox(0,0)[r]{0.4}}
\put(540,589){\makebox(0,0)[r]{0.2}}
\put(540,251){\makebox(0,0)[r]{0}}
\end{picture}
\caption{Wavefunctions in momentum space, $\chi(k)$ 
for four different values of the Fermi momentum, 
$k_F=0.1$ fm$^{-1}$, $k_F=0.4$ fm$^{-1}$, $k_F=0.8$ fm$^{-1}$ and 
$k_F=1.2$ fm$^{-1}$. The wavefunctions peak at the 
corresponding value of the 
Fermi momentum.}
\end{figure}
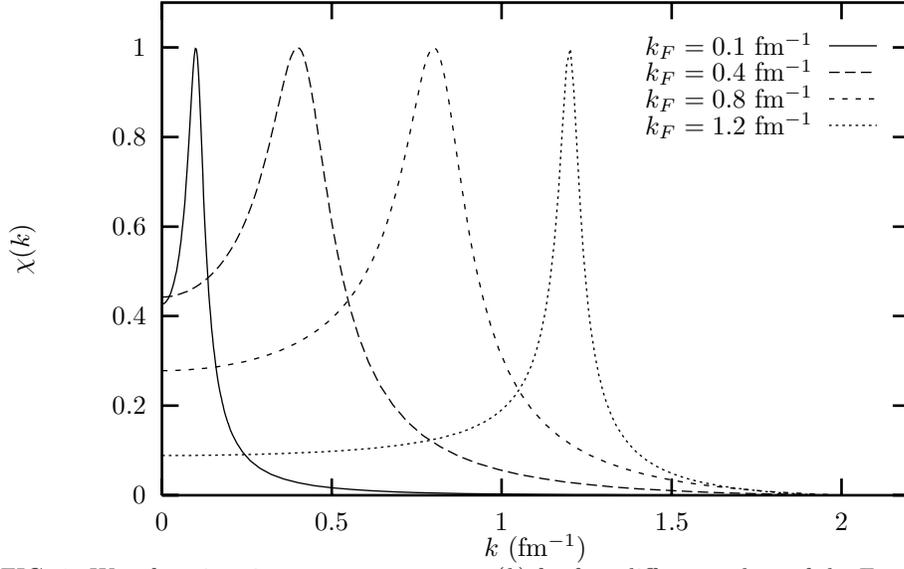

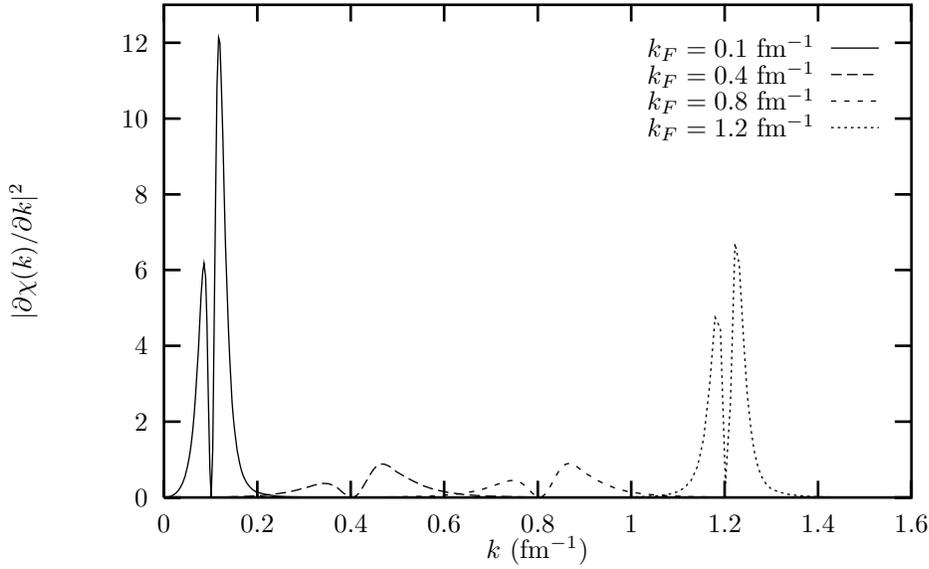
\begin{figure}
\setlength{\unitlength}{0.1bp}
\special{!
/gnudict 40 dict def
gnudict begin
/Color false def
/Solid false def
/gnulinewidth 5.000 def
/vshift -33 def
/dl {10 mul} def
/hpt 31.5 def
/vpt 31.5 def
/M {moveto} bind def
/L {lineto} bind def
/R {rmoveto} bind def
/V {rlineto} bind def
/vpt2 vpt 2 mul def
/hpt2 hpt 2 mul def
/Lshow { currentpoint stroke M
  0 vshift R show } def
/Rshow { currentpoint stroke M
  dup stringwidth pop neg vshift R show } def
/Cshow { currentpoint stroke M
  dup stringwidth pop -2 div vshift R show } def
/DL { Color {setrgbcolor Solid {pop []} if 0 setdash }
 {pop pop pop Solid {pop []} if 0 setdash} ifelse } def
/BL { stroke gnulinewidth 2 mul setlinewidth } def
/AL { stroke gnulinewidth 2 div setlinewidth } def
/PL { stroke gnulinewidth setlinewidth } def
/LTb { BL [] 0 0 0 DL } def
/LTa { AL [1 dl 2 dl] 0 setdash 0 0 0 setrgbcolor } def
/LT0 { PL [] 0 1 0 DL } def
/LT1 { PL [4 dl 2 dl] 0 0 1 DL } def
/LT2 { PL [2 dl 3 dl] 1 0 0 DL } def
/LT3 { PL [1 dl 1.5 dl] 1 0 1 DL } def
/LT4 { PL [5 dl 2 dl 1 dl 2 dl] 0 1 1 DL } def
/LT5 { PL [4 dl 3 dl 1 dl 3 dl] 1 1 0 DL } def
/LT6 { PL [2 dl 2 dl 2 dl 4 dl] 0 0 0 DL } def
/LT7 { PL [2 dl 2 dl 2 dl 2 dl 2 dl 4 dl] 1 0.3 0 DL } def
/LT8 { PL [2 dl 2 dl 2 dl 2 dl 2 dl 2 dl 2 dl 4 dl] 0.5 0.5 0.5 DL } def
/P { stroke [] 0 setdash
  currentlinewidth 2 div sub M
  0 currentlinewidth V stroke } def
/D { stroke [] 0 setdash 2 copy vpt add M
  hpt neg vpt neg V hpt vpt neg V
  hpt vpt V hpt neg vpt V closepath stroke
  P } def
/A { stroke [] 0 setdash vpt sub M 0 vpt2 V
  currentpoint stroke M
  hpt neg vpt neg R hpt2 0 V stroke
  } def
/B { stroke [] 0 setdash 2 copy exch hpt sub exch vpt add M
  0 vpt2 neg V hpt2 0 V 0 vpt2 V
  hpt2 neg 0 V closepath stroke
  P } def
/C { stroke [] 0 setdash exch hpt sub exch vpt add M
  hpt2 vpt2 neg V currentpoint stroke M
  hpt2 neg 0 R hpt2 vpt2 V stroke } def
/T { stroke [] 0 setdash 2 copy vpt 1.12 mul add M
  hpt neg vpt -1.62 mul V
  hpt 2 mul 0 V
  hpt neg vpt 1.62 mul V closepath stroke
  P  } def
/S { 2 copy A C} def
end
}
\begin{picture}(3600,2160)(0,0)
\special{"
gnudict begin
gsave
50 50 translate
0.100 0.100 scale
0 setgray
/Helvetica findfont 100 scalefont setfont
newpath
-500.000000 -500.000000 translate
LTa
600 251 M
2817 0 V
600 251 M
0 1858 V
LTb
600 251 M
63 0 V
2754 0 R
-63 0 V
600 537 M
63 0 V
2754 0 R
-63 0 V
600 823 M
63 0 V
2754 0 R
-63 0 V
600 1109 M
63 0 V
2754 0 R
-63 0 V
600 1394 M
63 0 V
2754 0 R
-63 0 V
600 1680 M
63 0 V
2754 0 R
-63 0 V
600 1966 M
63 0 V
2754 0 R
-63 0 V
600 251 M
0 63 V
0 1795 R
0 -63 V
952 251 M
0 63 V
0 1795 R
0 -63 V
1304 251 M
0 63 V
0 1795 R
0 -63 V
1656 251 M
0 63 V
0 1795 R
0 -63 V
2009 251 M
0 63 V
0 1795 R
0 -63 V
2361 251 M
0 63 V
0 1795 R
0 -63 V
2713 251 M
0 63 V
0 1795 R
0 -63 V
3065 251 M
0 63 V
0 1795 R
0 -63 V
3417 251 M
0 63 V
0 1795 R
0 -63 V
600 251 M
2817 0 V
0 1858 V
-2817 0 V
600 251 L
LT0
3114 1946 M
180 0 V
600 251 M
1 0 V
1 0 V
1 0 V
1 0 V
1 0 V
2 0 V
1 0 V
2 1 V
2 0 V
2 0 V
2 1 V
2 0 V
3 1 V
3 1 V
2 1 V
3 1 V
3 2 V
4 2 V
3 2 V
4 3 V
3 3 V
4 4 V
4 4 V
4 6 V
4 6 V
4 8 V
5 10 V
4 11 V
5 14 V
4 17 V
5 20 V
5 26 V
5 30 V
5 38 V
5 46 V
5 57 V
5 68 V
5 81 V
6 94 V
5 103 V
5 102 V
6 85 V
5 35 V
6 -51 V
5 -169 V
5 -273 V
6 -277 V
5 -112 V
6 184 V
5 448 V
6 524 V
5 399 V
6 178 V
5 -20 V
5 -148 V
6 -205 V
5 -214 V
5 -197 V
6 -170 V
5 -142 V
5 -115 V
5 -93 V
4 -74 V
5 -60 V
5 -48 V
5 -39 V
4 -32 V
5 -25 V
4 -21 V
4 -17 V
4 -15 V
4 -12 V
4 -10 V
4 -8 V
4 -8 V
3 -6 V
4 -5 V
3 -4 V
3 -4 V
3 -3 V
3 -3 V
2 -3 V
3 -2 V
2 -2 V
3 -2 V
2 -1 V
2 -1 V
1 -1 V
2 -1 V
1 -1 V
1 -1 V
1 0 V
1 -1 V
1 0 V
1 0 V
0 -1 V
3 0 V
12 -3 V
21 -5 V
31 -5 V
40 -3 V
49 -1 V
59 -1 V
66 -1 V
76 0 V
83 0 V
91 0 V
99 0 V
106 0 V
113 0 V
118 0 V
125 0 V
130 0 V
135 0 V
139 0 V
143 0 V
145 0 V
149 0 V
151 0 V
152 0 V
153 0 V
77 0 V
LT1
3114 1846 M
180 0 V
601 251 M
1 0 V
1 0 V
2 0 V
2 0 V
3 0 V
3 0 V
3 0 V
4 0 V
5 0 V
4 0 V
6 0 V
5 0 V
6 0 V
6 0 V
7 0 V
7 0 V
7 0 V
8 0 V
8 0 V
9 0 V
9 0 V
9 0 V
9 0 V
10 1 V
10 0 V
10 0 V
11 0 V
11 0 V
11 0 V
11 0 V
12 0 V
12 1 V
12 0 V
12 0 V
13 1 V
12 0 V
13 0 V
13 1 V
13 0 V
13 1 V
14 1 V
13 0 V
13 1 V
14 1 V
14 2 V
13 1 V
14 2 V
14 2 V
13 2 V
14 2 V
14 3 V
14 3 V
13 3 V
14 4 V
13 3 V
14 4 V
13 4 V
13 4 V
13 3 V
13 3 V
13 0 V
13 -1 V
12 -4 V
12 -6 V
13 -8 V
11 -10 V
12 -10 V
11 -8 V
11 -5 V
11 0 V
11 5 V
10 10 V
10 14 V
10 17 V
10 17 V
9 16 V
9 14 V
8 12 V
8 9 V
8 6 V
7 3 V
7 2 V
7 0 V
6 -1 V
6 -3 V
6 -2 V
5 -3 V
5 -3 V
4 -3 V
4 -3 V
3 -3 V
4 -2 V
2 -2 V
2 -2 V
2 -2 V
2 -1 V
1 -1 V
3 -1 V
10 -6 V
19 -11 V
28 -15 V
36 -17 V
44 -15 V
52 -11 V
59 -8 V
68 -5 V
74 -2 V
82 -2 V
88 -1 V
94 0 V
101 -1 V
106 0 V
111 0 V
116 0 V
120 0 V
125 0 V
127 0 V
130 0 V
133 0 V
135 0 V
77 0 V
LT2
3114 1746 M
180 0 V
601 251 M
2 0 V
2 0 V
4 0 V
4 0 V
5 0 V
5 0 V
6 0 V
7 0 V
8 0 V
9 0 V
9 0 V
10 0 V
11 0 V
11 0 V
12 0 V
13 0 V
13 0 V
14 0 V
15 0 V
15 0 V
16 0 V
16 0 V
18 0 V
17 0 V
18 0 V
19 0 V
19 0 V
20 0 V
20 0 V
20 0 V
21 0 V
22 0 V
21 0 V
22 0 V
23 0 V
23 0 V
23 0 V
23 0 V
23 0 V
24 0 V
24 0 V
24 0 V
25 1 V
24 0 V
25 0 V
24 0 V
25 0 V
25 0 V
25 0 V
24 1 V
25 0 V
25 0 V
24 1 V
25 0 V
24 1 V
24 0 V
24 1 V
24 1 V
24 2 V
23 2 V
23 2 V
23 2 V
22 4 V
22 4 V
22 4 V
21 6 V
21 7 V
21 7 V
20 7 V
20 6 V
19 5 V
18 0 V
19 -6 V
17 -12 V
17 -17 V
17 -17 V
16 -10 V
15 0 V
14 12 V
14 21 V
14 25 V
13 24 V
12 20 V
11 13 V
11 8 V
10 3 V
9 0 V
8 -3 V
8 -4 V
7 -4 V
6 -4 V
6 -5 V
5 -4 V
4 -3 V
3 -3 V
3 -2 V
2 -2 V
0 -1 V
3 2 V
9 -8 V
16 -9 V
23 -14 V
30 -15 V
37 -14 V
43 -12 V
50 -8 V
57 -6 V
63 -3 V
68 -3 V
74 -1 V
79 -1 V
85 0 V
89 0 V
93 -1 V
98 0 V
101 0 V
104 0 V
107 0 V
4 0 V
LT3
3114 1646 M
180 0 V
602 251 M
2 0 V
4 0 V
5 0 V
5 0 V
7 0 V
8 0 V
10 0 V
10 0 V
11 0 V
12 0 V
14 0 V
14 0 V
16 0 V
16 0 V
17 0 V
19 0 V
19 0 V
20 0 V
21 0 V
22 0 V
23 0 V
24 0 V
25 0 V
25 0 V
26 0 V
27 0 V
28 0 V
28 0 V
29 0 V
30 0 V
30 0 V
31 0 V
31 0 V
32 0 V
32 0 V
33 0 V
34 0 V
33 0 V
34 0 V
35 0 V
34 0 V
35 0 V
35 0 V
35 0 V
36 0 V
36 0 V
35 0 V
36 0 V
36 0 V
36 0 V
35 0 V
36 0 V
35 0 V
36 0 V
35 0 V
35 0 V
34 0 V
35 0 V
34 0 V
33 0 V
34 0 V
32 0 V
33 1 V
32 0 V
31 0 V
31 0 V
30 1 V
30 1 V
29 1 V
28 2 V
28 3 V
27 5 V
26 8 V
25 15 V
25 28 V
24 53 V
23 105 V
22 196 V
21 260 V
20 -41 V
19 -584 V
19 305 V
17 600 V
16 -80 V
16 -276 V
14 -212 V
14 -132 V
12 -81 V
11 -49 V
10 -32 V
9 -21 V
8 -14 V
7 -10 V
6 -7 V
5 -5 V
4 -3 V
2 -3 V
2 -1 V
2 -1 V
7 -3 V
13 -5 V
18 -6 V
25 -6 V
30 -4 V
35 -3 V
41 -2 V
45 -1 V
51 -1 V
55 0 V
60 -1 V
65 0 V
68 0 V
14 0 V
stroke
grestore
end
showpage
}
\put(3054,1646){\makebox(0,0)[r]{$k_F=1.2$ fm$^{-1}$}}
\put(3054,1746){\makebox(0,0)[r]{$k_F=0.8$ fm$^{-1}$}}
\put(3054,1846){\makebox(0,0)[r]{$k_F=0.4$ fm$^{-1}$}}
\put(3054,1946){\makebox(0,0)[r]{$k_F=0.1$ fm$^{-1}$}}
\put(2008,51){\makebox(0,0){$k$ (fm$^{-1}$)}}
\put(100,1180){%
\special{ps: gsave currentpoint currentpoint translate
270 rotate neg exch neg exch translate}%
\makebox(0,0)[b]{\shortstack{$|\partial\chi (k)/\partial k|^2$}}%
\special{ps: currentpoint grestore moveto}%
}
\put(3417,151){\makebox(0,0){1.6}}
\put(3065,151){\makebox(0,0){1.4}}
\put(2713,151){\makebox(0,0){1.2}}
\put(2361,151){\makebox(0,0){1}}
\put(2009,151){\makebox(0,0){0.8}}
\put(1656,151){\makebox(0,0){0.6}}
\put(1304,151){\makebox(0,0){0.4}}
\put(952,151){\makebox(0,0){0.2}}
\put(600,151){\makebox(0,0){0}}
\put(540,1966){\makebox(0,0)[r]{12}}
\put(540,1680){\makebox(0,0)[r]{10}}
\put(540,1394){\makebox(0,0)[r]{8}}
\put(540,1109){\makebox(0,0)[r]{6}}
\put(540,823){\makebox(0,0)[r]{4}}
\put(540,537){\makebox(0,0)[r]{2}}
\put(540,251){\makebox(0,0)[r]{0}}
\end{picture}
\caption{ Plot of $ |\partial \chi(k) / \partial k|^{2} $
proportional to the spread in space of the Cooper pair.
Legend as in Fig.\ 1.  }
\end{figure}

\begin{figure}
\input{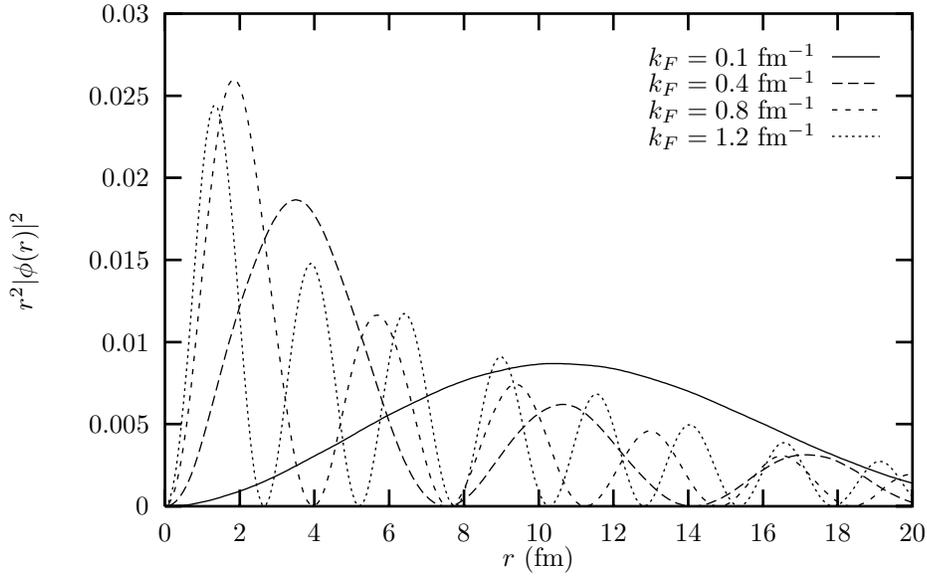}
\caption{$r^2|\phi (r)|^2$ as function of $r$.}
\end{figure}

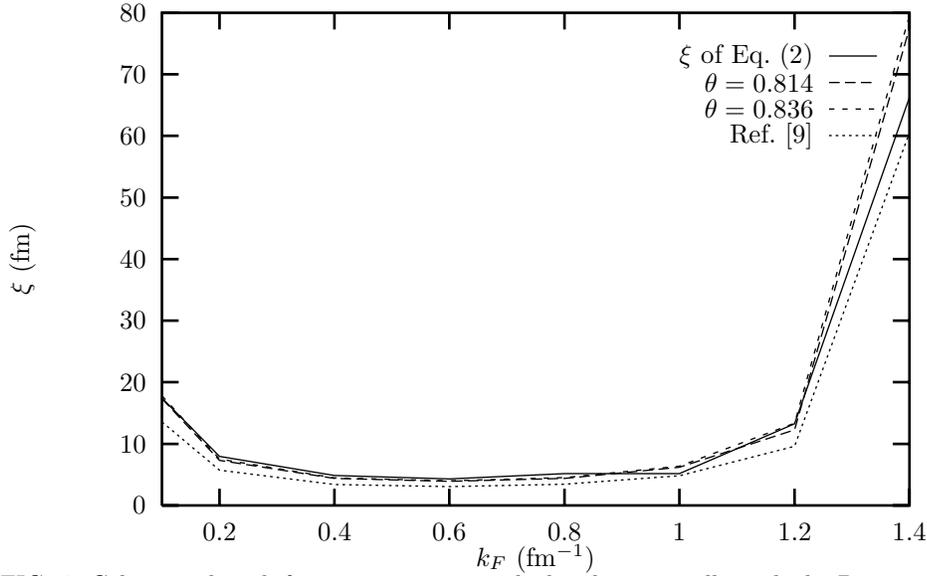
\begin{figure}
\setlength{\unitlength}{0.1bp}
\special{!
/gnudict 40 dict def
gnudict begin
/Color false def
/Solid false def
/gnulinewidth 5.000 def
/vshift -33 def
/dl {10 mul} def
/hpt 31.5 def
/vpt 31.5 def
/M {moveto} bind def
/L {lineto} bind def
/R {rmoveto} bind def
/V {rlineto} bind def
/vpt2 vpt 2 mul def
/hpt2 hpt 2 mul def
/Lshow { currentpoint stroke M
  0 vshift R show } def
/Rshow { currentpoint stroke M
  dup stringwidth pop neg vshift R show } def
/Cshow { currentpoint stroke M
  dup stringwidth pop -2 div vshift R show } def
/DL { Color {setrgbcolor Solid {pop []} if 0 setdash }
 {pop pop pop Solid {pop []} if 0 setdash} ifelse } def
/BL { stroke gnulinewidth 2 mul setlinewidth } def
/AL { stroke gnulinewidth 2 div setlinewidth } def
/PL { stroke gnulinewidth setlinewidth } def
/LTb { BL [] 0 0 0 DL } def
/LTa { AL [1 dl 2 dl] 0 setdash 0 0 0 setrgbcolor } def
/LT0 { PL [] 0 1 0 DL } def
/LT1 { PL [4 dl 2 dl] 0 0 1 DL } def
/LT2 { PL [2 dl 3 dl] 1 0 0 DL } def
/LT3 { PL [1 dl 1.5 dl] 1 0 1 DL } def
/LT4 { PL [5 dl 2 dl 1 dl 2 dl] 0 1 1 DL } def
/LT5 { PL [4 dl 3 dl 1 dl 3 dl] 1 1 0 DL } def
/LT6 { PL [2 dl 2 dl 2 dl 4 dl] 0 0 0 DL } def
/LT7 { PL [2 dl 2 dl 2 dl 2 dl 2 dl 4 dl] 1 0.3 0 DL } def
/LT8 { PL [2 dl 2 dl 2 dl 2 dl 2 dl 2 dl 2 dl 4 dl] 0.5 0.5 0.5 DL } def
/P { stroke [] 0 setdash
  currentlinewidth 2 div sub M
  0 currentlinewidth V stroke } def
/D { stroke [] 0 setdash 2 copy vpt add M
  hpt neg vpt neg V hpt vpt neg V
  hpt vpt V hpt neg vpt V closepath stroke
  P } def
/A { stroke [] 0 setdash vpt sub M 0 vpt2 V
  currentpoint stroke M
  hpt neg vpt neg R hpt2 0 V stroke
  } def
/B { stroke [] 0 setdash 2 copy exch hpt sub exch vpt add M
  0 vpt2 neg V hpt2 0 V 0 vpt2 V
  hpt2 neg 0 V closepath stroke
  P } def
/C { stroke [] 0 setdash exch hpt sub exch vpt add M
  hpt2 vpt2 neg V currentpoint stroke M
  hpt2 neg 0 R hpt2 vpt2 V stroke } def
/T { stroke [] 0 setdash 2 copy vpt 1.12 mul add M
  hpt neg vpt -1.62 mul V
  hpt 2 mul 0 V
  hpt neg vpt 1.62 mul V closepath stroke
  P  } def
/S { 2 copy A C} def
end
}
\begin{picture}(3600,2160)(0,0)
\special{"
gnudict begin
gsave
50 50 translate
0.100 0.100 scale
0 setgray
/Helvetica findfont 100 scalefont setfont
newpath
-500.000000 -500.000000 translate
LTa
600 251 M
2817 0 V
LTb
600 251 M
63 0 V
2754 0 R
-63 0 V
600 483 M
63 0 V
2754 0 R
-63 0 V
600 716 M
63 0 V
2754 0 R
-63 0 V
600 948 M
63 0 V
2754 0 R
-63 0 V
600 1180 M
63 0 V
2754 0 R
-63 0 V
600 1412 M
63 0 V
2754 0 R
-63 0 V
600 1645 M
63 0 V
2754 0 R
-63 0 V
600 1877 M
63 0 V
2754 0 R
-63 0 V
600 2109 M
63 0 V
2754 0 R
-63 0 V
817 251 M
0 63 V
0 1795 R
0 -63 V
1250 251 M
0 63 V
0 1795 R
0 -63 V
1683 251 M
0 63 V
0 1795 R
0 -63 V
2117 251 M
0 63 V
0 1795 R
0 -63 V
2550 251 M
0 63 V
0 1795 R
0 -63 V
2984 251 M
0 63 V
0 1795 R
0 -63 V
3417 251 M
0 63 V
0 1795 R
0 -63 V
600 251 M
2817 0 V
0 1858 V
-2817 0 V
600 251 L
LT0
3114 1946 M
180 0 V
600 656 M
817 436 L
433 -72 V
433 -13 V
434 20 V
433 0 V
434 188 V
433 1231 V
LT1
3114 1846 M
180 0 V
600 654 M
817 421 L
433 -68 V
433 -11 V
434 11 V
433 42 V
434 141 V
433 1515 V
LT2
3114 1746 M
180 0 V
600 665 M
817 426 L
433 -71 V
433 -11 V
434 12 V
433 43 V
434 162 V
433 1538 V
LT3
3114 1646 M
180 0 V
600 566 M
817 384 L
433 -54 V
433 -8 V
434 9 V
433 32 V
434 111 V
433 1183 V
stroke
grestore
end
showpage
}
\put(3054,1646){\makebox(0,0)[r]{Ref.\ [9]}}
\put(3054,1746){\makebox(0,0)[r]{$\theta=0.836$}}
\put(3054,1846){\makebox(0,0)[r]{$\theta=0.814$}}
\put(3054,1946){\makebox(0,0)[r]{$\xi$ of Eq.\ (2)}}
\put(2008,51){\makebox(0,0){$k_F$ (fm$^{-1}$)}}
\put(100,1180){%
\special{ps: gsave currentpoint currentpoint translate
270 rotate neg exch neg exch translate}%
\makebox(0,0)[b]{\shortstack{$\xi$ (fm)}}%
\special{ps: currentpoint grestore moveto}%
}
\put(3417,151){\makebox(0,0){1.4}}
\put(2984,151){\makebox(0,0){1.2}}
\put(2550,151){\makebox(0,0){1}}
\put(2117,151){\makebox(0,0){0.8}}
\put(1683,151){\makebox(0,0){0.6}}
\put(1250,151){\makebox(0,0){0.4}}
\put(817,151){\makebox(0,0){0.2}}
\put(540,2109){\makebox(0,0)[r]{80}}
\put(540,1877){\makebox(0,0)[r]{70}}
\put(540,1645){\makebox(0,0)[r]{60}}
\put(540,1412){\makebox(0,0)[r]{50}}
\put(540,1180){\makebox(0,0)[r]{40}}
\put(540,948){\makebox(0,0)[r]{30}}
\put(540,716){\makebox(0,0)[r]{20}}
\put(540,483){\makebox(0,0)[r]{10}}
\put(540,251){\makebox(0,0)[r]{0}}
\end{picture}
\caption{Coherence length for neutron matter
calculated numerically with  
the Bonn potential (solid line) 
using Eqs.\ (2)-(4). The  line with long dashes 
are for a $\xi$ which 
fits the numerical points with $\theta=0.814$  while for the short-dashed line 
$\theta=0.836$, see text for further details. The dotted line represents
$\xi$ given by an approximation frequently used in the vortex pinning 
literature [9].}
\end{figure}

\end{document}